# Pressure dependence of the monoclinic phase in (1-x)Pb(Mg$_{1/3}$Nb$_{2/3}$)O$_3$-xPbTiO$_3$ solid solutions

Muhtar Ahart[1], Stanislav Sinogeikin[2], Olga Shebanova[2], Daijo Ikuta[2], Zuo-Guang Ye[3], Ho-kwang Mao[1], R. E. Cohen[1], and Russell J. Hemley[1]

[1]Geophysical Laboratory, Carnegie Institution of Washington, 5251 Broad Branch Road, Washington, D.C. 20015, USA

[2]HPCAT, Geophysical Laboratory, Carnegie Institution of Washington, Argonne, Illinois 60439, USA

[3]Department of Chemistry and 4D LABS, Simon Fraser University, Burnaby, BC, V5A 1S6, Canada

We combine high-pressure x-ray diffraction, high-pressure Raman scattering, and optical microscope to investigate a series of (1-x)Pb(Mg$_{1/3}$Nb$_{2/3}$)O$_3$-xPbTiO$_3$ (PMN-xPT) solid solutions (x=0.2, 0.3, 0.33, 0.35, 0.37, 0.4) in diamond anvil cells up to 20 GPa at 300 K. The Raman spectra show a peak centered at 380 cm$^{-1}$ starting above 6 GPa for all samples, in agreement with previous observations. X-ray diffraction measurements are consistent with this spectral change indicating a structural phase transition; we find that the triplet at the pseudocubic (220) Bragg peak merges into a doublet above 6 GPa. Our results indicate that the morphotropic phase boundary region (x=0.33 to 0.37) with the presence of monoclinic symmetry persists up to 7 GPa. The pressure dependence of ferroelectric domains in PMN-0.32PT single crystals was observed using a polarizing optical microscope. The domain wall density decreases with pressure and the domains disappear at a modest pressure of 3 GPa. We propose a pressure – composition phase diagram for PMN-xPT solid solutions.

## Introduction

Lead-based relaxor ferroelectrics with complex perovskite structures exhibit a strong frequency-dispersive dielectric susceptibility [1, 2]. Their high piezoelectric constants make them suitable for applications in devices such as those used in sonar or medical imaging [3]. Extensive theoretical and experimental studies [4-9] have greatly advanced our understanding of relaxors. Many details remain to be elucidated due to the complexity of the materials, which have a high degree of compositional, structural, and polar disorder.

Single crystals of the solid solutions between relaxors and ferroelectrics, namely $(1-x)$Pb(Mg$_{1/3}$Nb$_{2/3}$)O$_3$-$x$PbTiO$_3$, known as PMN-$x$PT, are being used now as a new generation of electromechanical devices. When properly oriented, they have piezoelectric coefficients and electromechanical coupling factors which are the highest yet reported, with strain level one order of magnitude larger than that of conventional piezoelectric PbZrO$_3$-PbTiO$_3$ (PZT) ceramics. The exceptional electromechanical properties of these relaxor-based solid solutions are related to the phase boundary (in the temperature–composition phase diagram) between the rhombohedral and the tetragonal phases, the so-called morphotropic phase boundary (MPB), which is a common feature of many ferroelectric-based solid solutions. From the temperature-composition phase diagram [8, 9], the phase sequence of PMN-xPT at room temperature can be stated as follows: rhombohedral R3m phase for x<0.32, intermediate monoclinic Cm phase for 0.32<x<0.37, and tetragonal P4mm phase for x>0.37.

Application of pressure can tune the physical properties of relaxors and introduce new phenomena, as observed by several researchers [10-14]. Samara [10] pointed out that pressure is an important parameter in the investigations of relaxor-to-ferroelectric crossover phenomena in disordered perovskite systems. A high pressure Raman scattering study of compositional dependencies of PMN-xPT system [11] led to the establishment of its preliminary pressure-composition phase diagram, which indicates that the low PT compositions at low pressure maintain their rhombohedral symmetry up to 6 ~ 7 GPa, while the higher pressure phases are non-cubic. Motivated by the strong interest in understanding the pressure effects on these highly disordered ferroelectrics, we employed a combination of high pressure Raman scattering, x-ray diffraction and optical microscopic observation methods to investigate the phase behavior of the PMN-xPT solid solutions under pressure at 300 K. Our purpose is twofold: (i) to investigate the pressure induced phase transition, and the pressure dependence of domain structure in PMN-32PT, and (ii) to establish a pressure-composition phase diagram [14].

## Experiments

Powder samples of the PMN-xPT solid solutions were prepared by solid-state reaction according to the process described in Ref. 8. Single crystals of PMN-0.32PT were obtained from TRS Inc. A crystal with dimensions of 70×70×20 μm$^3$ and orientated with the largest face parallel to (111)$_{cub}$ was loaded into a diamond anvil cell (DAC) with a 4:1 mixture of methanol and ethanol as the pressure medium for domain observations under pressure. A chip of ruby was also loaded for pressure determination. Small amounts of powder samples of different composition of PMN-xPT solid solutions were loaded into separate DACs with neon as the pressure medium for x-ray diffraction and Raman scattering experiments.

X-ray diffraction experiments were carried out at Beamline 16-ID-B and 16-BM-D of HPCAT at the Advanced Photon Sources (Argonne National Laboratory). In this experiment, a pre-monochromator combined with a Si(311) crystal was employed to provide the monochromatic (24 keV) incident beam with a wavelength close to 0.4 Å. A MAR3450 image plate was used to record oscillation photographs. Standard ruby luminescence technique was used to determine the pressure, with an estimated accuracy of 0.2 GPa. Additional experimental details about Beamline-16 (HPCAT) can be found in Ref. 15.

It should be noted that these three techniques observe different length scales; the optical microscope we used has one micron resolution, therefore we could not observe nanosized polar domains. However we were able to observe the macro (size is larger than one micron) domains' behavior under pressure. Raman scattering uses a long wave length laser as its excitation source, giving optical phonons near the Γ point of the Brillouin zone; and considering our beam size at focal point, we are collecting signal from a couple micron sized regions, but the smallest regions that interact with incident light will be in the order of tens of nm due to the laser wave length, thus we observe the dynamics that occur at the nm sized local structures. The x-ray diffraction probes the regions that are comparable to the size of unit cell of crystals considering the x-ray wavelength. However, it is difficult to resolve the distribution of $Mg^{2+}$, $Nb^{5+}$, and $Ti^{4+}$ cations on the B-site of the perovskite structure in these solid solutions. The short range ordering would appear as the diffuse scattering around Bragg peaks in these materials in the single crystal x-ray diffractions as shown in many studies. Here, we perform the powder diffractions, thus we will not consider the effect of diffuse scattering in our study.

## Results and Discussion
### (1) Domain observations

We used a polarized optical microscope to observe the pressure dependence of domain structure of PMN-0.32PT up to 8 GPa. Figure 1 shows the observed domain pattern as a function of pressure. The rhombohedral domain pattern is visible at modest pressure 0.6 GPa, with the domain walls forming the characteristic 109 degree or 71 degrees angles. The domain density decreases or disappears with pressure. No domains are visible above 5 GPa.

### (2) **Raman scattering**

Fig. 2 shows a set of representative Raman spectra of PMN−xPT (x=0.22 and 0.37) at selected pressures. The bands are broad in comparison to the first-order scattering of conventional ferroelectrics, such as $PbTiO_3$, which show sharp peaks in their polar phases. We employed a single grating system as our spectrometer for Raman scattering with a higher throughput, which is not sensitive to the low frequency modes centered around 50 $cm^{-1}$. Previous work [11-13] has shown that the mode at 50 $cm^{-1}$ in lead-based relaxor ferroelectrics has a slight pressure dependence, hardening on compression, and does not show any drastic change with temperature and pressure. Here we concentrate on two particular regions in the Raman spectra. The first is region around 350 $cm^{-1}$, a sharp peak appears above 4 to 7 GPa for all samples and increases in intensity with increasing pressure. The second region is that around 550 $cm^{-1}$, where there are two broad peaks that merge together above 5-7 GPa for all samples. To obtain additional information, we normalized the measured spectra with the Bose-Einstein factor n(ω)+1, where $n(\omega)=[(\exp(\hbar\omega/k_BT)-1]^{-1}$, $\omega$ is frequency, $\hbar$ is Planck's constant, $k_B$ is the Boltzmann constant and T is the temperature. We then decomposed the measured profiles using a multipeak fitting procedure. Reasonable fits were obtained under the assumption that peaks are described by spectral functions of damped harmonic oscillators. Although the correct method for decomposing the Raman spectra of relaxor ferroelectrics is still under debate due to their complexity [16-18], here we used the minimum number of peaks needed to obtain a good fit to the spectra. Our results reflected the general trend of the Raman spectra of these materials under pressure. As an example, we show the pressure dependencies of the Raman bands of PMN-0.37PT in Fig. 3.

Although PMN, an end member of PMN-xPT solid solutions, has an average cubic perovskite structure, it shows a Raman spectrum similar to that of the other compositions. It is believed that the observed strong Raman scattering in PMN arises from local chemically ordered regions [17]. Experimental evidence for locally ordered regions has been obtained from transmission electron microscopy (TEM) which indicates chemical short-range ordered domains are embedded in a disordered matrix [19]. Thus, microstructure is characterized by a 1:1 chemical order parameter on the 50 Å length scale, and not by composition fluctuations about Mg:Nb=1:2. Specifically, 1:1 ordered regions could have NaCl type Mg/Nb ordering with one B-site occupied by $Nb^{5+}$ (we call it: *a*), and the other by a random mixture of $1/3Nb^{5+} + 2/3Mg^{2+}$ [19, 20] (we call it: *b*), therefore B-site of these relaxors forms ordering as *ababab* ... on the 50 Å length scale and which borders with the chemically disordered regions throughout the sample. Thus, these 1:1 ordered regions would have Fm-3m structure. The main features of the Raman spectrum of PMN reflect the Fm-3m (normal Raman mode: $2F_{2g}+E_g+A_{1g}$; for details please see Ref. 21) symmetry and are consistent with rocksalt ordering. Although PMN-PT solid solutions differ from PMN, the Raman spectra of these solid solutions are similar to that of PMN. Thus, we apply this simple model (normal Raman mode: $2F_{2g}+E_g+A_{1g}$) to describe observed Raman spectra and based on previously established systematics for the perovskites [20, 21], we assign the band at 780 $cm^{-1}$ to the $A_{1g}$ mode and the broad band at 500-600 $cm^{-1}$ to the $E_g$ mode. The intense band at 50 $cm^{-1}$ and the broad band at 272 $cm^{-1}$ we assigned to the $F_{2g}$ modes. The remaining Raman features, (such as, the Raman peak around 430 $cm^{-1}$, and the band around 780 $cm^{-1}$ shows more than three peaks for higher PT content samples), are more complex and cannot be explained simply using this picture; they are evidence for lowering of symmetry from the Fm-3m selection rules. The polar and chemical disorder cause vibration modes to have coherence lengths that are small compared with their wavelengths. Such modes are not characterized by a single wavevector and will not obey momentum selection rules; all wavevectors contribute to the scattering process, causing broad features [22, 23].

The $A_{1g}$ band at 780 $cm^{-1}$ results from the mode of fully symmetrical breathing vibration of oxygen octahedra. As shown in Figs. 2 and 3, this mode slightly hardens with pressure, its linewidth decreases only slightly, and it is insensitive to both temperature and pressure [12]. The $E_g$ band at 550 $cm^{-1}$ corresponds to the antiphase breathing mode of oxygen octahedra. This mode is degenerate and splits into two modes under a tetragonal, rhombohedral or monoclinic distortion [20, 21].

As one of the most remarkable results, we find that pressure suppresses the local distortions. This can be seen in the two peaks assigned as $E_g$ merge together with pressure and becomes a diffuse peak at 5 GPa (Figs. 2 and 3). Similar behavior is observed for PT under pressure [24]. In PT, for example, the peaks between 500-600 cm$^{-1}$ merges together with pressure around 10 GPa, thus one can consider this effect, a reduced local distortion, somehow relate to the reduced ferroelectric behavior with pressure and the formation of a paraelectric-like phase. In PMN-xPT, the $F_{2g}$ bands at 50 and 272 cm$^{-1}$ all slightly harden with pressure. Also, a peak at 350 cm$^{-1}$, which overlaps with the broad band at 272 cm$^{-1}$, increases its intensity on compression. This effect can be explained by the pressure-induced tilting of octahedra or by the appearance of the $F_{2g}$ vibrations at high pressure due to the lowering of the potential barriers between the ferroelectricity-related wells. The appearance of the $F_{2g}$ vibration at high pressure is consistent with our explanation for the $E_g$ mode. The pressure dependencies of the Raman spectrum indicate that the polar correlations decrease with increasing pressure. This general conclusion is in good agreement with the observations by Samara et al. [10].

**(3) X-ray diffraction**

Diffraction data were collected directly from the powder samples within the diamond anvil cell using an image plate detector. Although the samples were oscillated from -6º to 6º in omega to improve powder averaging, incorrect intensity ratios can occur if significant preferred orientation is present in the samples. However, the peak positions and hence the derived unit cell parameters are not affected by this problem. The pressure dependence of the diffraction patterns were measured between ambient pressure and 20 GPa. The six pseudocubic reflections of (100), (110), (111), (200), (220), and (222), are needed to determine the crystal symmetry within the limits of the instrumental resolution, together with the corresponding lattice parameters. For the data analysis, we followed the procedure proposed by Noheda et al. [8] that the individual reflection profiles were fitted to a pseudo-Voigt function appropriately corrected for asymmetry with intensity, peak position, full width at half maximum (FWHM) and mixing parameter as variables. When appropriate, the peak widths and mixing parameters were constrained to be the same for all the peaks in a set of data. Additional peaks were added based on the difference patterns, the relative intensities, and the goodness-of-fit residuals. Compared to the standard Rietveld-type technique of profile-matching, in which the full diffraction pattern is fitted with an assumed structural model, the advantage of this procedure is that it does not require any assumptions about the phases to be included. We have found that anisotropic broadening of linewidth is a common effect in Pb-based perovskite systems, and is likely to occur when one or more lattice parameters is more sensitive to compositional fluctuations than the others, and are consistent with previous observation [8].

The pressure evolution of the diffraction patterns for PMN-0.22PT, 0.33PT and 0.40PT is shown in Fig. 3. These three compositions are located to the left, middle and right sides, respectively, of the MPB in the PMN-xPT phase diagram [8]. The diffraction profiles around the pseudo-cubic (111) and (220) diffraction peaks from PMN-0.22PT, 0.33PT and 0.40PT at selected pressures are shown in Figs. 3a, 3b and 3d, respectively, in which the least-squares fits to the data points are shown as solid curves with the de-convoluted peaks. In each case, the observed diffraction peaks show no evidence of phase coexistence. The pseudo-cubic (111) peak splits to two for PMN-0.22PT, indicating that PMN-0.22PT has rhombohedral symmetry; such splitting exists even at higher pressure (8.9 GPa). When combined with the results of Raman scattering, we conclude that PMN-0.22PT undergoes a rhombohedral to rhombohedral phase transition. The pseudo-cubic (111) peak splits into three or more for PMN-0.33PT, indicating a monoclinic distortion. With increasing pressure, the triplet splitting of the pseudo-cubic (111) peak merges to two above 6 GPa, indicating a monoclinic to rhombohedral phase transition in PMN-0.33PT. It is interesting that the PMN-0.40PT sample behaves differently. This material is tetragonal at ambient pressure, as seen in the doublet splitting of the pseudo-cubic (220) peak and it undergoes a tetragonal to monoclinic transition at a modest pressure of 2 GPa. Based on our results we have mapped out the pressure-composition phase diagram for the PMN-xPT solid solutions at room temperature, as shown in Fig. 4. The phase boundary points for PT and PMN are taken from Refs. 24 and 25. It should be noted that the high-pressure rhombohedral phase could be *R*-3*c*. The evidences are coming from our results and previous first-principle calculations [26]. Our domain observations show no domains are visible above 5 GPa indicating a possible center-symmetric space group for the high-pressure phase. Our diffraction data point to a rhombohedral structure for all studied samples above 6-7 GPa. Furthermore, first-principle calculation [26] proposed R-3c is a stable phase for a similar system. Thus we propose a rhombohedral *R*-3*c* for the high pressure phase of PMN-xPT solid solutions.

The presence of a morphotropic phase boundary (MPB) with monoclinic symmetry [8] seems to be a common feature for the relaxor-based solid solutions like PMN-xPT, where their electromechanical properties are maximal. The MPB region persists up to 7 GPa in the PMN-xPT system. Pressure suppresses the polar nanoregions in relaxor materials. For example, several single crystal x-ray diffraction experiments [25, 27, 28] observed characteristic diffuse scattering around Bragg peaks in these relaxors and disappearance of diffuse scattering at a modest pressure (around 4 GPa). These studies also suggest that the correlations between polar nanoregions in relaxors are the main contributor to the diffuse scattering and are consistent with the theoretical calculations [28]. Therefore pressure suppresses the correlations between polar nanoregions in relaxos. Thus our results indicate that polar nanoregions, which are believed to be responsible for many intriguing properties of relaxors, are not necessary for the existence of an MPB region in the phase diagram. This result is consistent with first-principle calculations [29]. In fact, Refs. 6 and 7 show that even a single compound, lead titanate, displays a MPB region under pressure. Our results also reveal that complex microstructure or compositions might not be necessary to obtain high coupling piezoelectricity and that the transition in complex relaxor ferroelectrics or high performance ferroelectric solid solutions with lead titanate as an end member can be tuned to take place down to zero pressure by changing composition. These results indirectly confirm the suggestion made in Ref. 7 that chemical pressures are important in these solid solutions, for example, PMN-0.40PT is a tetragonal at ambient pressure and undergoes a transition into a monoclinic phase at a modest pressure of 2 GPa. This can be interpreted as the substitution of 60% of PMN for PT has tuned the MPB transition pressure in PT from a higher pressure to around 2 GPa and the transition temperature from a lower temperature to room temperature by chemical pressure.

The equation of state (EOS) is obtained from high pressure data, usually with diffraction data, and it describes the relationships among the thermodynamic variables: volume (V), pressure (P), temperature (T), and energy (E). It is therefore an expression of structure and chemical bonding at the atomic level. The application of pressure is of prime importance for the EOS studies because pressure varies the interatomic spacing, resulting in large changes in structure, bonding, and electronic configuration. In our case, pressure driven structural transitions are observed in these solid solutions, for example, peak splitting; and we also can use our diffraction data to calculate the pressure-volume relationships.

The Vinet function [30] was used to fit the pressure-volume data:

$$P = \frac{3K_0(1-X)}{X^2}\exp[\frac{3}{2}(K_0' - 1)(1 - X)],$$

where $P$ is the pressure, $X=(V/V_0)^{1/3}$ is the relative volume, $K_0$ is the bulk modulus, and $K_0$' is its pressure derivative. Figure 5 shows the pressure dependencies of the volumes. We fit the Vinet EOS to the data and obtain $V_0$=62.7 ± 0.3 Å$^3$, $K_0$=100 ± 10 GPa, and $K_0$'=4 (fixed) for PT; $V_0$=65.6 ± 0.3 Å$^3$, $K_0$=123 ± 10 GPa, and $K_0$'=4 (fixed) for PMN-0.33PT; $V_0$=64.7 ± 0.2 Å$^3$, $K_0$=120 ± 9 GPa, and $K_0$'=4 (fixed) for PMN-0.22PT. The EOS data for PMN, $V_0$=66.33 ± 0.03 Å$^3$, $K_0$=104 ± 5 GPa, and $K_0$'=4.7 ± 2 are adapted from Ref. 27. The bulk moduli are typical for Pb based relaxors. These results also indicate that the bulk moduli and initial volumes have composition dependencies for PMN-xPT solid solutions. Both bulk moduli and initial volume show maxima near the morphotropic phase boundary region. Similar behaviors are also observed for piezoelectric and elastic anisotropic properties in PMN-PT solid solutions (Ref. 31) and PZN-PT solid solutions (Ref. 32). In these solid solutions, piezoelectric and elastic anisotropic properties show strong composition dependence and exhibit maxima near the MPB boundary from rhombohedral side in the composition-temperature phase diagrams. The origin of the high piezoelectric response in relaxor ferroelectric materials is attributed to the ease of rotating the polarization from the rhombohedral to tetragonal orientation by the application of an electric field in the [001] direction in these materials via the MPB boundary [5]. First principle calculations of PbTiO$_3$ under pressure identified a tetragonal-to-monoclinic-to-rhombohedral-to-cubic phase transition sequence and a colossal enhancement of the piezoelectric response near the phase transitions [6]. This work showed that the existence of the monoclinic phase is not unique to solid solution ferroelectrics. The only necessary condition is that the high symmetry phases are very close in energy and therefore rotation from one phase to the other can occur through the low-symmetry monoclinic phase. In the DFT work on high pressure PbTiO$_3$ they find that the $C_{44}$ elastic constant goes to zero as the tetragonal to monoclinic phase transition is approached and the $C_{11}$-$C_{12}$ (in cubic axis) goes to zero at the rhombohedral to monoclinic phase transition. This result for PbTiO$_3$ matches the observations of an enhanced bulk modulus and mechanical anisotropy on the rhombohedral side near the MPB (i.e. $C_{11}$-$C_{12}$ goes to zero therefore bulk modulus increases) for PMN-*x*PT. The microstructure of PMN-*x*PT is still under much debate, and extrinsic contributions due to domains may also influence the electromechanical behavior for these materials. However, our observations suggest the extrinsic contributions from polar nanoregions are minimal to their phase diagram in these solid solutions.

**Conclusions**

We performed high-pressure x-ray diffraction, high-pressure Raman scattering, and optical microscopy experiments to study a series of PMN-*x*PT solid solutions (x=0.2, 0.3, 0.33, 0.35, 0.37 and 0.4) in diamond anvil cells up to 20 GPa at 300 K. The Raman spectra show a new peak centered at 380 cm$^{-1}$ above 6 GPa for all samples, consistent with previous observations. X-ray diffraction results are in agreement with this spectral change, indicating a structural phase transition. We find that the triplet at the pseudocubic (220) Bragg peak merges into a doublet above 6 GPa. The results indicate that the morphotropic phase boundary (x=0.33 to 0.37) with a monoclinic symmetry persists up to 7 GPa. The pressure dependence of ferroelectric domains in PMN-0.32PT single crystals was observed with a polarizing optical microscope. The domain densities decrease with pressure and disappear at a modest pressure of 3 GPa.

**Acknowledgement**

This work was sponsored by the Office of Naval Research under Grants No. N00014-02-1-0506, No. N00014-06-1-0166 and No. N00014-11-1-0552; the Carnegie/Department of Energy Alliance Center (CDAC, DE-FC03-03NA00144). Use of the Advanced Photon Source was supported by the U. S. Department of Energy under Contract No. DE-AC02-06CH11357. HPCAT operations are supported by CIW, CDAC, UNLV and LLNL through funding from DOE-NNSA and DOE-BES, with partial instrumentation funding by NSF.

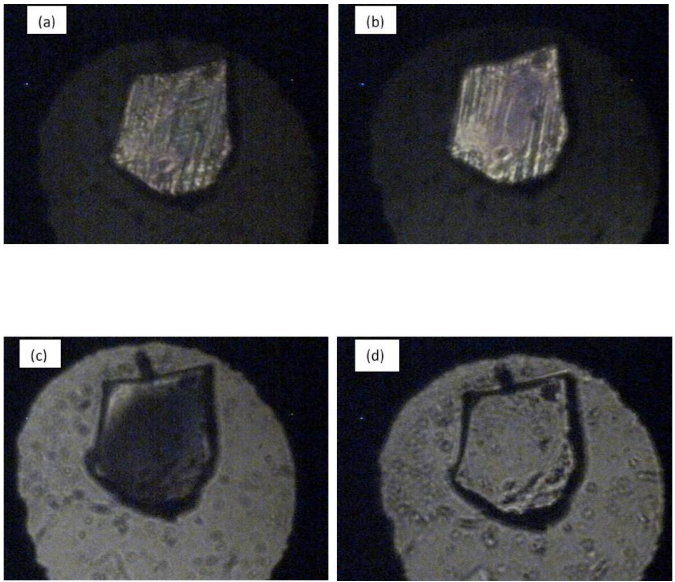

Figure 1. Observed domains in a (111)$_{cub}$ PMN-0.32PT crystal at selected pressures. (a) 0.3 GPa; (b) 1.4 GPa; (c) 3 GPa; and (d) 6 GPa. Domains disappear above 3 GPa. The Lateral size of sample is about 80 μm.

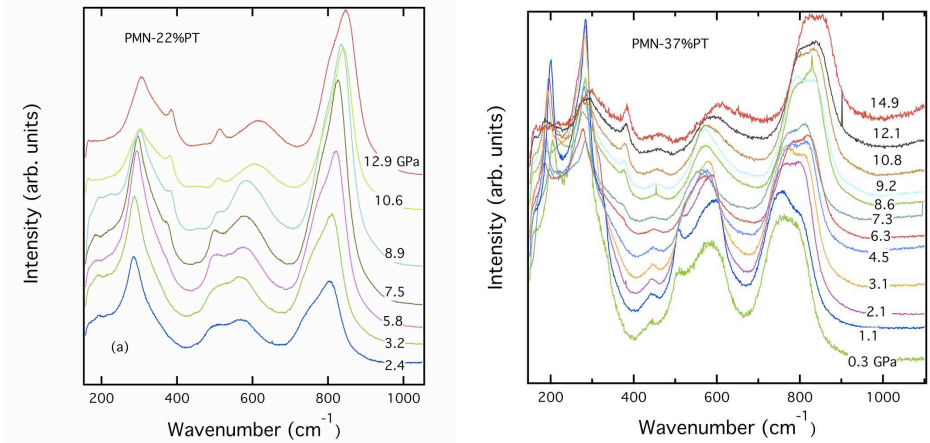

Figure 2. Examples of Raman spectra for (a) PMN-0.22PT; (b) PMN-0.37PT. The main features of the Raman spectrum of PMN-xPT reflect the Fm-3m symmetry. A new peak splits from broad $F_{2g}$ band at 380 cm$^{-1}$, which can be explained by the pressure-induced tilting of oxygen octahedra.

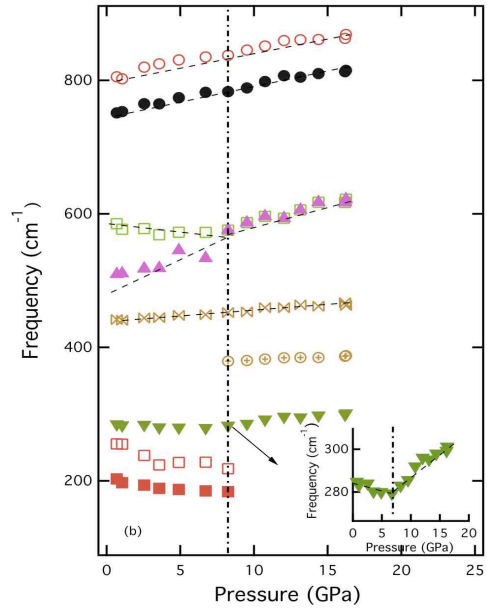

Figure 3. Pressure dependencies of Raman bands of PMN-0.37PT. A new peak appears near the 370 cm$^{-1}$. We use a monochromatic spectrometer with the notch filters to cut off the signal below 150 cm$^{-1}$ to avoid strong Rayleigh scattering. Thus it is difficult to resolve the lower frequency modes at higher pressure, as shown in the figure. Marks are data and dashed curves are guides for eyes.

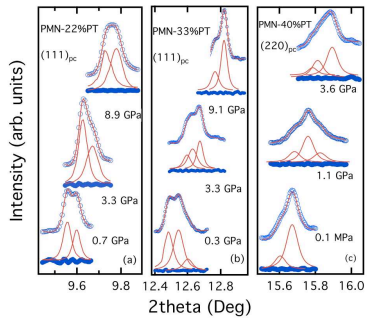

Figure 4. The profiles for the diffraction patterns for selected samples at selected pressures. Open circles are the experimental data, solid curves are the least square fits to the data and solid circles are the residuals. (a) Pressure dependence of the pseudo-cubic (111) peak from the PMN-0.22PT sample; (b) pressure evolution of the pseudo-cubic (111) peak from the PMN-0.33PT sample; (c) pressure dependence of the pseudo-cubic (220) peak from the PMN-0.40PT sample.

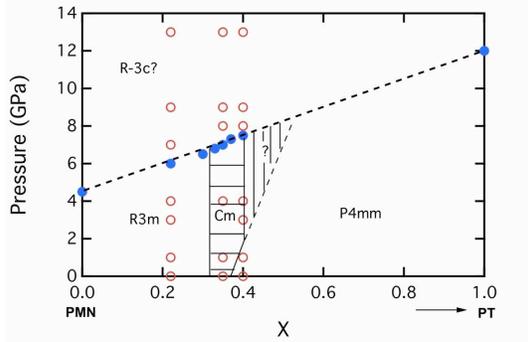

Figure 5. Pressure-composition phase diagram of PMN-xPT solid solutions at room temperature. Marks are data. The solid circles represent the structural phase transition boundary with pressure. The vertical lines represent the MPB region with a monoclinic symmetry. The phase boundary points for PMN and PT are taken from Refs. 11 and 24. Although we cannot determine high-pressure phase (structure) rigorously, however we propose possible high-pressure phase is *R*-3c based on our measurements.

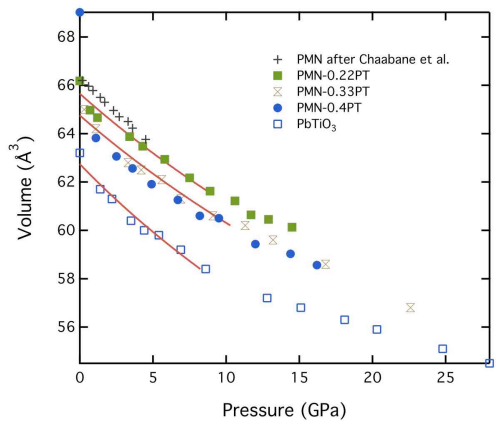

Fig. 6. Pressure dependencies of the volumes for PT, PMN-0.4PT, PMN-0.33PT, PMN-0.22PT, and PMN. Marks are experimental data, curves are fits (Vinet EOS) and the data for PMN sample are adapted from Ref. 26.